\documentstyle[prc,aps]{revtex}

\begin{document}
\draft


\title{On the choice of colliding beams to study deformation effects
in relativistic heavy ion collisions}

\author{S. Das Gupta and C. Gale}

\address{
Physics Department, McGill University\\ 
3600 University St., Montr{\'e}al, Qu{\'e}bec \\ Canada H3A 2T8\\ }

\date{ \today }

\maketitle

\begin{abstract}
It has been suggested that collisions between deformed shapes
will lead to interesting effects on various observables such as $K$
production and elliptic flow.  Simple formul{\ae} can be written down
which show how to choose the colliding beams which will maximise
the effects of deformation.  
\end{abstract}





In the nuclear periodic table there are large regions where nuclei
are deformed with prolate intrinsic shapes.  Recent calculations 
\cite{shuryak,li} suggest that if one can arrange to have
tip-on-tip collisions (collision axis = symmetry axis) of two prolate 
shapes, the physical results are
significantly different from when the collisions are side-on-side
(the collision axis is perpendicular to the symmetry axis).  
If unpolarised beams are used, interesting events will be buried
among many uninteresting ones.  The occurrence of the interesting 
events can be enhanced by using polarised beams.  We do a quantitative 
estimate here.

For unpolarised beam, the average density is spherical.  One still
will see effects of deformation because we shall assume that heavy 
ion collisions
sample many-body correlations contained in  
$|\Psi(\vec r_1,\vec r_2,....\vec r_A)|^2$: the collision knocks out
all particles.  Having a particle at $\vec r_1$ influences the
probability of seeing a particle at $\vec r_2$ etc.  But not having a 
control of the overall orientation is a problem: we would, for
example, like to enhance the chance of tip-on-tip collision.  It is
then necessary to control even the one-body density.


We use the rotational model \cite {BM} to extract answers.  The 
wavefunction of
the ground state of an odd-A deformed nucleus can be written as
\begin{eqnarray}
\Psi_{IMK}=\sqrt{\frac{2I+1}{16\pi^2}}\left[{D_{MK}^I}^{*}(\Omega)\, 
\Phi_K(x')+
(-)^{I-K} {D_{M-K}^I}^{*} (\Omega) \, \Phi_{-K}(x')\right]\ .
\end{eqnarray}
The symmetrisation will play no role in what follows so we will use
\begin{eqnarray}
\Psi_{IMK}=\sqrt{\frac{2I+1}{8\pi^2}}{D_{M,K}^I}^{*} (\Omega) \Phi_K(x')\ .
\end{eqnarray}
We use the convention of Rose \cite {Rose} for $D$ functions.  The quantity
$\Phi_K(x')$ consists of two parts:
\begin{eqnarray}
\Phi_K(x')=\Phi_0(x')\phi_k(x')\ .
\end{eqnarray}
The $\Phi_0(x')$ is the intrinsic deformed state of the
even-even core and $\phi_k(x')$ is the Nilsson
model type orbital: $\phi_k(x')=\sum_jc_{jk}|jk>$.  One may wish to
include antisymmetrisation between the core and the extra nucleon
but for what we do later this will not matter.  For many applications
$\Phi_0(x')$ plays no role and is suppressed.  As usual, $\Omega\equiv
\alpha,\beta,\gamma$ are the Eulerian angles specifying the orientation
of the deformed intrinsic state with respect to the lab.  All of 
this is, of course, very standard rotational model.

We need to choose $I,M$ such that the one-body density in the lab is as
deformed as possible.  This is not so transparent.  However, if the
expectation value of the quadrupole moment is large the density should
show large deformation since the quadrupole mode is the basic deviation
from sphericity.  To evaluate the expectation value of the quadrupole
operator $r^2Y_{20}(x)$ in the lab we express the operator in the 
body-fixed system: $r^2Y_{20}(x)=\sum_m {D_{0 m}^2}^{*} (\Omega) 
\, r^2Y_{2m}(x')$.
One then obtains
\begin{eqnarray}
<\Psi_{I M K}|r^2Y_{20}(x)|\Psi_{I M K}>=(I2M0|IM)(I2K0|IK) \times
<\Phi_K(x')|r^2Y_{20}(x')|\Phi_K(x')>
\end{eqnarray}
In this equation, the first two terms on the right hand side are
Clebsch-Gordan coefficients, the third one is the deformation of the
intrinsic state.  One might want to argue that the last orbital is
just one of many orbitals and can be dropped: 
$<\Phi_K(x')|r^2Y_{20}(x')
|\Phi_K(x')>\, \approx\, <\Phi_0(x')|r^2Y_{20}(x')|\Phi_0(x')>$.  In that case
the only role of the last odd particle is to align the nucleus.  The
product of the two Clebsch-Gordan coefficients is the reduction in
perfect alignment brought about by quantum mechanics.
 
Since we are in the ground state, $K=I$.  Hence the key factor is
\begin{eqnarray}
R=(I2M0|IM)(I2I0|II)
\end{eqnarray}
Clearly for $M=I$ one has approximate alignment in the 
direction of the symmetry axis.  The value of $(I2I0|II)^2$ is
$\frac{I(2I-1)}{(I+1)(2I+3)}$.  This goes to 1 as $I\rightarrow\infty$.
This is the limit at which the frequency of tip-on-tip collision
is one hundred percent.

For odd-A nucleus, assuming the direction of z is defined by the collision
axis, we need to have $|I,1/2>$ to have the ``best'' body-on-body
collisions.  The reduction factor is simply calculated by the above
formula.

The arguments presented here for odd-A nuclei should hold for odd-odd
nuclei also.  It is advantageous to choose a nucleus with large ground
state spin.  Ground state spins 9/2 in the deformed regions are available.
Perhaps the most advantageous nucleus from this point view is $^{176}$Lu
which has 7 as its ground state spin.

If a density in the intrinsic state $\rho_d(x')$ is assumed, the 
one-body density
of the state $\Psi_{I M K}$ can be numerically computed from
$\frac{2I+1}{4\pi}\int|d_{M,K}^I(\beta)|^2sin\beta\, d\beta\, d\gamma\, 
\rho_d(x',\beta, \gamma)$.  

To conclude, we find that in order to study the role of 
deformation in high energy
heavy ion collisions it will be judicious to choose nuclei with high
spin in the ground state.  By choosing $|II>$ states where the z-axis
is the collision axis one can enhance tip-on-tip
collision.  By choosing $|I\, 1/2>$ states one can enhance
body-on-body collisions.  Quantum mechanics will prevent a perfect alignment.

 
We are thankful to E. Shuryak for emphasizing to us the many-body
correlation aspects of relativistic heavy ion collisions, and to J.
Barrette, C. M. Ko, and C. Pruneau for useful exchanges.
This work is supported in part by the Natural Sciences
and Engineering Research Council of Canada and in part by the Fonds FCAR
of the Qu\'ebec Government.


\begin{references}
\bibitem{shuryak}E. Shuryak, Phys. Rev. C 61, 034905 (2000).
\bibitem{li}Bao-An Li, Phys. Rev. C 61, 021903 (2000).
\bibitem{BM} A. Bohr and B. R. Mottelson, {\it Nuclear Structure, Vol.II}
(W.A. Benjamin, Reading, 1975).

\bibitem{Rose} M. E. Rose, {\it Elementary Theory of Angular Momentum}
(John Wiley and Sons Inc, New York, 1957).


\end{references}
\end{document}